\documentclass[twocolumn,aps,pre]{revtex4}
\usepackage{amsmath}
\usepackage{graphicx}
\usepackage{bm}
\usepackage{epsfig}

\makeatletter

\begin{document}
\title{Study of quantum Otto heat engine using driven-dissipative  Schr\"{o}dinger
equation}
\author{You-wei Fang,$^{1}$ Yu-ting Zheng$^{1}$ and Jun Chang$^{1,}$}
\email{junchang@snnu.edu.cn}
\address{\textsuperscript{1}College of Physics and Information Technology,
Shaanxi Normal University, Xi'an 710119, China}

\date{\today}
\begin{abstract}
The quantum heat engines have drawn much attention due to miniaturization of devices recently. We study the dynamics of the quantum
Otto heat engine using the driven-dissipative Schr\"{o}dinger equation. Starting
from different initial states, we simulate the time evolutions of the
internal energy, power and heat-work conversion
efficiency. The initial state impacts on these thermodynamic quantities
before the Otto cycle reaches stable. In the transition period, the
efficiency and power may be higher or lower than the corresponding
values in the cyclostationary state. Remarkably, the efficiency could
surpass the Otto limit and even the Carnot limit and the power could be much higher than the rated power. The  efficiency anomaly is due to the energy in the initial state. Thus, we suggest that periodically pumping could take the similar role of a hot bath but could be manipulated flexibly. Furthermore, we propose a new quantum engine
working in a single reservoir to convert the pump energy into mechanical
work. This manipulative engine could potentially be applied to working in the microenvironments
without a large temperature difference, such as the biological tissues
in vivo. Our protocol is expected to model a new quantum engine with
the advantage of applicability and controllability. 
\end{abstract}
\maketitle

\section{introduction}

Converting heat into mechanical work periodically, the heat engines
as the generators of motion triggered both the industrial revolution
and the theoretical development of thermodynamics in the $18$th and
$19$th centuries. Recently, the heat engines have been re-invigorated
by miniaturization of devices  \cite{Peter2009}. Experimentally, elaborate
efforts have been made on the design of the machines employing quantum
systems, such as harmonic oscillators  \cite{Lin2003}, ultracold atoms  \cite{Brantut2013},
a single particle, such as an atom, ion, electron \cite{Abah2012,Koski2014,Shubhashis2014,Ronagel2016,von2019}
and molecular \cite{Hill2005} as well as a spin-1/2 system \cite{Geva_1992,George2011,von2019,Peterson2019}
and qubit system \cite{Noah2010,Nicolas2012,Bohr_Brask_2015}. They
work at the atom level to convert heat or light into mechanical or
electrical power \cite{Benenti2017,Jan2017}. Theoretically, various
model systems performing thermodynamic cycles have been proposed as
the quantum heat engines (QHEs), dating back to the three-level masers
proposal in 1959 \cite{Scovil1959}. Recently, more attentions are
drawn to the relationship between different types of QHEs \cite{Quan2007,Uzdin2015,Cyril2017,Klatzow2019},
irreversible work and inner friction \cite{Rezek_2006,Plastina2014},
the impact of coherence \cite{Scully2003,Rahav2012,Goswami2013}
and the application prospects and the transition to refrigerators \cite{Gelbwaser-Klimovsky2013,Kosloff2014,Uzdin_2014,Patrick2016,Camati2020}.
These QHEs not only comply with the classical laws of thermodynamics
but also follow the rules of quantum mechanics \cite{Quan2005,Kosloff2013,Paul2014}.
Consequently, the quantum thermodynamics is boosted by engineering
a variety of quantum devices.

Among the QHEs, the quantum Otto heat engine (QOHE), mimicking the
common four-strokes car engine, has been extensively studied \cite{Tova2003,Kieu2004,Hubner2014,Abah2014,Kosloff2017,Gentaro2017,Patrice2019,Das2020,Wiedmann_2020}.
The working substance is only a quantum harmonic oscillator with a
tunable vibration frequency. The heat engine is alternately coupled
to a hot and a cold reservoir. The $4$-stroke Otto cycle consists
of two isochoric processes and two adiabatic processes. During the
isochoric processes, the QOHE exchanges heat with the hot or cold
baths but keeps the frequency constant. In the adiabatic processes,
the work input or output changes the frequency of the oscillator.

In this paper, we take advantage of the driven-dissipative Schr\"{o}dinger
equation to study the time evolution of the quantum Otto cycle process
and evaluate the performance of QOHE. The time-dependent probability distribution
of the oscillator's energy levels is calculated by the driven-dissipative
Schr\"{o}dinger equation. The time evolutions of the internal energy,
power and efficiency are simulated. The quantum machine starts
from the ground state, the coherent state and the state with equal probability distribution
on several lower energy levels, respectively. In the first case,
the efficiency and power increase with the number of the Otto cycles
until they reach a stable value. In the latter two cases, the efficiency
could be larger than the Otto limit, or even the Carnot limit, and the power could be much higher than the rated power. The overrange is due to the contribution of the energy stored in the initial
state. It implies an effective method to manipulate the QOHE. Thus, we suggest that periodically pumping quantum heat
engine could enhance the rated power and also could substitute the role of a hot bath. Therefore, we propose a possible
design of a new quantum engine working in a single bath to convert
the pump energy into the mechanical work. Such an engine may be used
for working under control in the microenvironments. We also study the balance between
the net power output and the machine efficiency since it is impossible
to realize the maximums of both the efficiency and power output at the same time.

\section{Model of quantum Otto heat engine\label{sec:Model-of-quantum}}

We apply the driven-dissipative Schr\"{o}dinger equation \cite{Chang2010,Veenendaal2010}
to study the time evolution of the QOHE. The energy exchange between
the QOHE and the heat reservoirs is described by the driven-dissipative
operator $D$. Previously, we call it the dissipative operator in
the decaying process \cite{Chang2010,Veenendaal2010}. This equation
could deal with strong system-environment coupling and the substantial
environmental memory effects, as demonstrated in our previous work \cite{Chang2012}.
The dissipative Schr\"{o}dinger equation with the time-dependent
quantum state is written as 

\begin{equation}
i\hbar\frac{d\left|\psi\left(t\right)\right\rangle }{dt}=\left(H_{0}+iD\right)\left|\psi\left(t\right)\right\rangle ,
\end{equation}
where the Hamiltonian $H_{0}$ of the quantum harmonic oscillator
with a time-dependent frequency reads

\begin{equation}
H_{0}=\hbar\omega\left(t\right) a^{\dagger}a,\label{eqH}
\end{equation}
where $a^{\dagger}$ and $a$ are the Bosonic creation and annihilation
operators. Alternatively, 
\begin{equation}
H_{0}=\underset{n}{\sum}E_{n}\left|n\right\rangle \left\langle n\right|,\label{eqH0}
\end{equation}
where the energy of the ground state $\left|0\right\rangle $ is assumed
to be zero, $\left|n\right\rangle $ is the $n$-bosons state of the
oscillator, and the corresponding energy $E_{n}=n\hbar\omega$.

Selecting $\left|n\right\rangle$ as the basis, we write the system wave vector 
\begin{eqnarray}
| \psi(t)\rangle=\sum_n c_n(t)|n\rangle,
\end{eqnarray}
with the coefficient $c_n$ in terms of an amplitude $a_n(t)=|c_n(t)|$ and a phase $\varphi_n(t)$, or $c_n(t)=a_n(t) e^{i\varphi_n(t)}$. We can express the change in the coefficient due to the presence of the bath
\begin{eqnarray}
\left . \frac{dc_n(t)}{dt}\right |_B =
\left . \frac{da_n(t)}{dt}\right |_B  e^{i\varphi_n(t)}
+i \left . a_n(t) e^{i\varphi_n(t)} \frac{d{\varphi_n(t)}}{dt}\right |_B. 
\end{eqnarray}
In general, the coupling to the environment affects both the
probability and the phase of the system. The latter term gives the change in phase induced by its surroundings. Due to the complexity of the surroundings, its nature usually only is taken into account in an effective way. Here, we assume that the phase of the local system
is changed randomly by the large number of degrees of freedom of the
surroundings which results in a total phase change close to zero or a constant according to the law of large numbers. 
We therefore only consider the changes in the probability by the environment. Below we give the explicit expression for the change in probabilities $P_n(t)=a_n^2(t)$. The change in the coefficient due to the bath is then given by
\begin{eqnarray}
\left . \frac{dc_n(t)}{dt}\right |_B 
=\frac{1}{2a_n(t)}\left . \frac{dP_n(t)}{dt}\right |_B  e^{i\varphi_n(t)}  
=\left . \frac{1}{2}\frac{d\ln P_n(t)}{dt}\right |_B  c_n(t).
\end{eqnarray}
This leads to the $D$ operator describes the state changes of the system by surroundings,
which is given by
\begin{equation}
D=\frac{\hbar}{2}\underset{n}{\sum}\frac{d\ln P_{n}\left(t\right)}{dt}\left|n\right\rangle \left\langle n\right|.
\end{equation}
The time evolution of the probability distribution function $P_{n}\left(t\right)$
of the $n$th state is described by the rate equation \cite{Chang2010}
\begin{align}
\frac{dP_{n}\left(t\right)}{dt} & =-2n\Gamma P_{n}\left(t\right)+2\left(n+1\right)\Gamma P_{n+1}\left(t\right)\nonumber \\
 & \ -2\Gamma e^{-\hbar\omega/k_{B}T}\left[-nP_{n-1}\left(t\right)+\left(n+1\right)P_{n}\left(t\right)\right],\label{eqdP}
\end{align}
where $\Gamma=\Gamma_0 (n_{BE}+1)$ with the Bose-Einstein distribution $n_{BE}$ and the relaxation constant  $\Gamma_0$ of the oscillator induced
by the reservoirs. The factor $e^{-\hbar\omega/k_{B}T}$ is introduced
to ensure the rate equation obeying the usual detailed balance relations. The first two terms at the right side of Eq. (\ref{eqdP}) describe the change of $P_n$ due to emitting a boson to the bath, and the other two terms for the change of $P_n$ due to absorbing a boson from the bath. 

\begin{figure}

\includegraphics[width=0.8\columnwidth]{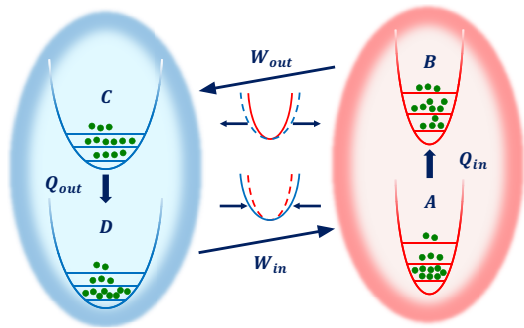}

\caption{Schematic cycle of the quantum Otto heat engine mimicking a classical
$4$-cycle engine. The cycle includes two isochoric processes ($A\rightarrow B,\ C\rightarrow D$)
and two adiabatic processes ($B\rightarrow C,\ D\rightarrow A$).
In the isochoric phases, the oscillator absorbs the heat $Q_{in}$
from the hot bath with the temperature $T_{h}$ and releases the heat
$Q_{out}$ to the cold bath with the temperature $T_{c}$, respectively.
In the adiabatic phases without entropy change, the work output $W_{out}$
and work input $W_{in}$ induce the decrease and increase of the oscillator
frequency, respectively. The horizontal line segments in the parabolic
curves represent the energy levels of the oscillator and the bottom horizontal
line indicates the ground state. \label{fig1}}
\end{figure}

Starting from a given initial population distribution $P_{n}(t=0),$
we solve the driven-dissipative Schr\"{o}dinger equation
and obtain the temporal variation of $P_{n}(t)$. Thus, the time-dependent
internal energy of the QOHE reads

\begin{equation}
U(t)=\underset{n}{\sum}E_{n}P_{n}(t).\label{eqU}
\end{equation}
The internal energy of the engine may be changed by the work output
or input as well as the exchange heat with a bath. According
to the first law of thermodynamics $\mkern3mu\mathchar'26\mkern-12mu dU=\mkern3mu\mathchar'26\mkern-12mu dQ+\mkern3mu\mathchar'26\mkern-12mu dW$,
then \cite{Kieu2004,Kieu2006}

\begin{equation}
\mkern3mu\mathchar'26\mkern-12mu dQ=\underset{n}{\sum}E_{n}dP_{n};\ \mkern3mu\mathchar'26\mkern-12mu dW=\underset{n}{\sum}P_{n}dE_{n}.\label{eqQW}
\end{equation}

During the thermal cycle, the occupation probability apportion $P_{n}$ of the $n$-bosons state changes in the endothermic and exothermic processes, and keeps constant in the adiabatic processes. If there are no coherence in the $\left|n\right\rangle$ states, the von Neumann entropy $S(t)$ as function
of occupation distribution $P_{n}(t)$ is written as 
\begin{equation}
S(t)=-k_{B}\underset{n}{\sum}P_{n}(t)lnP_{n}(t),\label{eqS}
\end{equation}
where $k_{B}$ is the Boltzmann constant.

Let address the $4$-stroke Otto cycle in more detail. In the isochoric
processes, the heat exchange with the heat reservoirs changes the
particle occupation allocation $P_{n}(t)$ without work done or $\mkern3mu\mathchar'26\mkern-12mu dW=0$,
i.e., in the stages $A\rightarrow B$ and $C\rightarrow D$, as shown
in Fig. \ref{fig1}. In the adiabatic or isentropic processes,
the oscillator frequency of the QHE changes slowly with time to ensure that there is no change of the particle occupation on each level or $dP_{n}=0$, and then the entropy $S(t)$ keeps a constant. There is also no heat
exchange with the heat reservoirs or $\mkern3mu\mathchar'26\mkern-12mu dQ=0$,
i.e., in the stages $B\rightarrow C$ and $D\rightarrow A$.

$A\rightarrow B$: starting from an initial state at the point $A$, the QOHE contacts
the hot bath with the temperature $T_{h}$. The oscillator is heated up by the hot bath and
keeps its frequency $\omega_{h}$ constant.  To the point $B$, it
stops exchanging heat with the hot bath. The total absorbed heat is 
\begin{equation}
Q_{in}=\underset{n}{\sum}n\hbar\omega_{h}\left(P_{n}^{B}-P_{n}^{A}\right).\label{eqQin}
\end{equation}

$B\rightarrow C$:  the machine does work in adiabatic expansion step. The von Neumann entropy remains unchanged, $dP_{n}=0$
or $P_{n}^{C}=P_{n}^{B}$. The oscillation frequency relaxes from
$\omega_{h}$ to $\omega_{c}$, and the work output is
\begin{equation}
W_{out}=\sum_{n}n\hbar\left(\omega_{h}P_{n}^{B}-\omega_{c}P_{n}^{C}\right).\label{eqWout}
\end{equation}

$C\rightarrow D$: the heat engine couples to the cold bath at the temperature $T_{c}$. The QOHE keeps the same frequency $\omega_{c}$
and releases the heat $Q_{out}$ to the cold bath with 
\begin{equation}
Q_{out}=\underset{n}{\sum}n\hbar\omega_{c}\left(P_{n}^{C}-P_{n}^{D}\right).\label{eqQout}
\end{equation}

$D\rightarrow A'$: the adiabatic compression starts and the frequency of the oscillator is enhanced from
$\omega_{c}$ to $\omega_{h}$ by the input work 
\begin{equation}
W_{in}=\sum_{n}n\hbar\left(\omega_{h}P_{n}^{A'}-\omega_{c}P_{n}^{D}\right),\label{eqWin}
\end{equation}
with $P_{n}^{A'}=P_{n}^{D}$. It is worthy of noting that the $P_{n}^{A'}$
is not always equal to the $P_{n}^{A}$ unless the Otto cycle is periodically
stable.

The effective or net work output $W_{eff}$ throughout the Otto cycle
is given by

\begin{equation}
W_{eff}=W_{out}-W_{in}.\label{eqWeff}
\end{equation}
Substituting the Eqs. \eqref{eqWout} and \eqref {eqWin} into Eq. \eqref{eqWeff}, the total effective work
$W_{eff}$ done per Otto cycle is written as
\begin{equation}
W_{eff}=\sum_{n}n\hbar\left[\left(\omega_{h}P_{n}^{B}-\omega_{c}P_{n}^{C}\right)-\left(\omega_{h}P_{n}^{A'}-\omega_{c}P_{n}^{D}\right)\right].\label{eqWe}
\end{equation}

Applying the conventional definition, the heat-work conversion efficiency
$\eta$ for the QOHE is given by

\begin{equation}
\eta=\frac{W_{eff}}{Q_{in}}.\label{eqeta0}
\end{equation}
Combining Eq. \eqref{eqWe} and \eqref{eqQin}, the efficiency $\eta$
is rewritten as
\begin{equation}
\eta=\frac{\underset{n}{\sum}n\hbar\left[\omega_{h}\left(P_{n}^{B}-P_{n}^{A'}\right)-\omega_{c}\left(P_{n}^{C}-P_{n}^{D}\right)\right]}{\underset{n}{\sum}n\hbar\omega_{h}\left(P_{n}^{B}-P_{n}^{A}\right)}.\label{eqeta}
\end{equation}
We assume the frequency of the oscillator changes slowly enough in the expansion and compression process with no population redistribution or without inner friction \cite{Rezek,Patrice2019,Plastina2014}. One has $P_{n}^{B}=P_{n}^{C},\ P_{n}^{D}=P_{n}^{A'}$.
Then, the efficiency $\eta$ can be obtained by simplifying Eq. \eqref{eqeta}
as

\begin{equation}
\eta=\eta_{O}\frac{\underset{n}{\sum}n\left(P_{n}^{B}-P_{n}^{A'}\right)}{\underset{n}{\sum}n\left(P_{n}^{B}-P_{n}^{A}\right)},\label{eqeta-1}
\end{equation}
with the Otto efficiency $\eta_{O}=1-\omega_{c}/\omega_{h}$. 

It is worthy of noting that in the efficiency definition, the contribution of energy  in the initial state is ignored. $P_{n}^{A}$ could be very different to $P_{n}^{A'}$ or the value in the stable cycles when the machine is away from a stable cycle, which could lead to the efficiency strongly deviated from the Otto limit. For example, starting from a high energy state, the machine even releases heat into the hot bath or $Q_{in} < 0$, and then the efficiency becomes negative. In particular, if we prepare the stating state with $P_{n}^{A}=P_{n}^{B}$, we even get the divergence of the efficiency. In the following, we show that if the energy in the initials state is taken into account, the maximum of the efficiency is still the Otto limit.

\section{Time evolution of quantum Otto heat engine\label{sec:3}}

\subsection*{A. Otto Cycle without formation of thermal balance \label{subsec:A.Non-Steady-State-QOHE}}

In this section, we simulate the time evolution process of the $4$-stroke
Otto cycle without formation of thermal balance with the baths. Each period includes four same time segments, e.g., $\tau=\tau_{AB}=\tau_{BC}=\tau_{CD}=\tau_{DA}$.

In all the numerical calculations, we set the lower oscillator energy $\hbar\omega_{c}$ as the
energy unit, and $\tau_{c}=2\pi/\omega_{c}$ as the time unit.

During the working stroke, the oscillator frequency reducing or increasing
is similar to the volume expanding or compressing in the classical
model. We set the time evolution of the frequency function as $\omega\left(t\right)=\omega_{h,c}-t(\omega_{h,c}-\omega_{c,h})/\tau$.
The harmonic frequency gradually decreases from $\omega_{h}$ to $\omega_{c}$
during the time $\tau_{BC}$ in the work output stages and gradually
increases from $\omega_{c}$ to $\omega_{h}$ during the time $\tau_{DA}$
in the work input stages. These working strokes take place under adiabatic
conditions, and no heat exchange with heat reservoirs or $\mkern3mu\mathchar'26\mkern-12mu dQ=0$. During the thermal exchanging stroke, the vibration frequency is
kept constant, and the oscillator exchanges heat with the reservoir
without work done or $\mkern3mu\mathchar'26\mkern-12mu dW=0$.

\begin{figure}
\includegraphics[width=0.6\columnwidth]{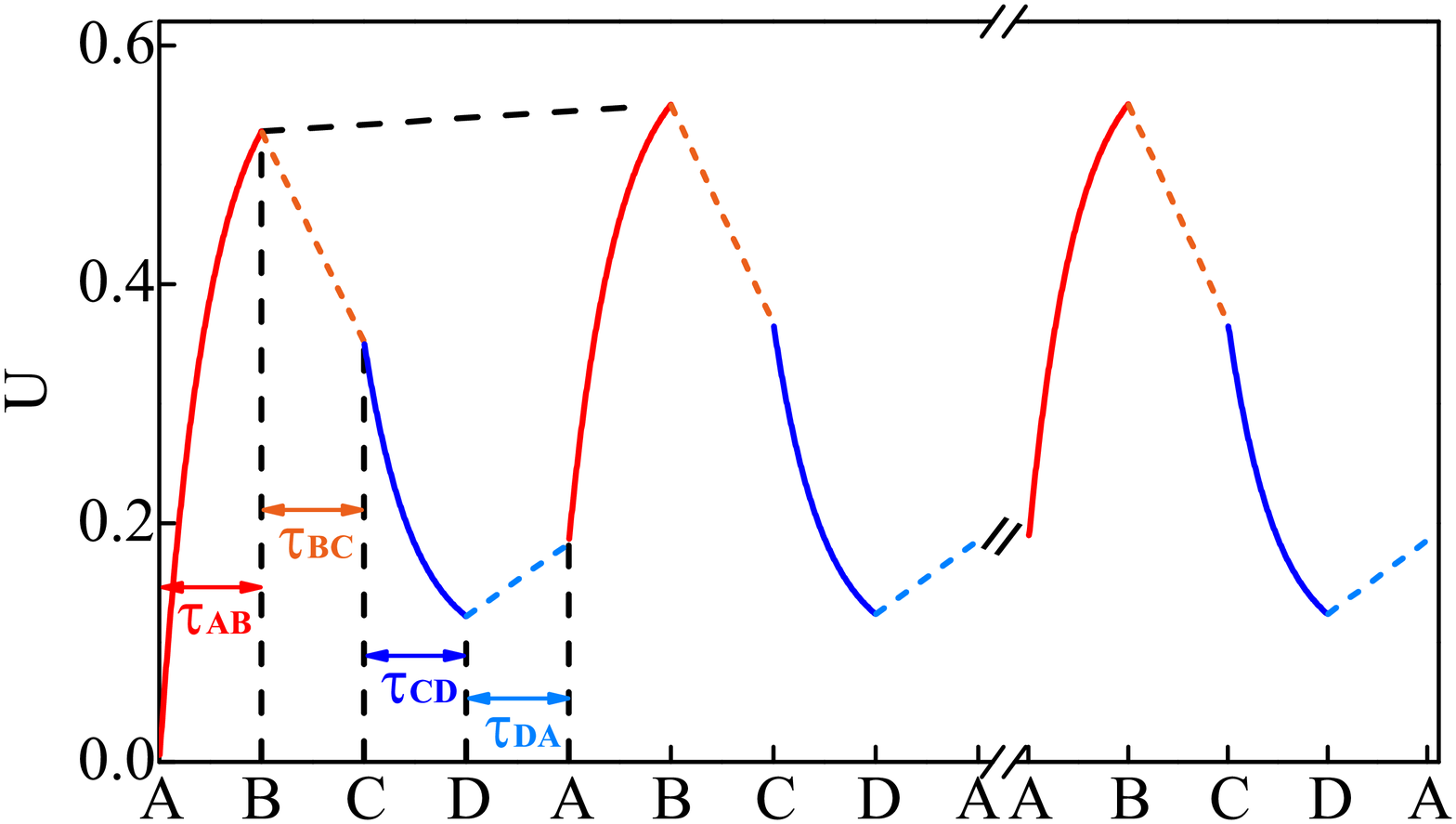}

\caption{The time evolutions of the internal energy $U$ in the QOHE starting from the ground state to the periodic steady
state. The red and blue solid lines indicate the two isochoric processes in the duration $\tau_{AB}$ and $\tau_{CD}$.
The orange and light-blue dashed lines indicate the two adiabatic
processes with the constant entropies in the segments $\tau_{BC}$ and $\tau_{DA}$. In our numerical calculations,
we set $\hbar\omega_{c}$ as the unit of energy, and $\tau_{c}=2\pi/\omega_{c}$
as the unit of time. The working duration $\tau=2.0$. The harmonic
oscillators of low frequency $\hbar\omega_{c}=1.0$ and high frequency
$\hbar\omega_{h}=1.5$. The temperature of cold bath $k_{B}T_{c}=0.4$
and the temperature of the hot bath $k_{B}T_{h}=1.2$, respectively.
Here, we define the relaxation time constant between the oscillator and the reservoirs
$\left(2\Gamma_0\right)^{-1}=1$. The right side of the separator shows the results when
the Otto cycle reaches a periodical stable state. The top black oblique
dashed line indicates that the difference of the internal energy $U$ at the points $B$ and $B'$ before the machine
reaches a stable state.
\label{fig2}}
\end{figure}

The time evolution of the internal energy $U$ and efficiency
$\eta$ depend on the time-dependent probabilities of the oscillator's energy levels. The
occupancy distribution is determined by solving the driven-dissipative Schr\"{o}dinger equation.

\begin{figure}
\includegraphics[width=0.6\columnwidth]{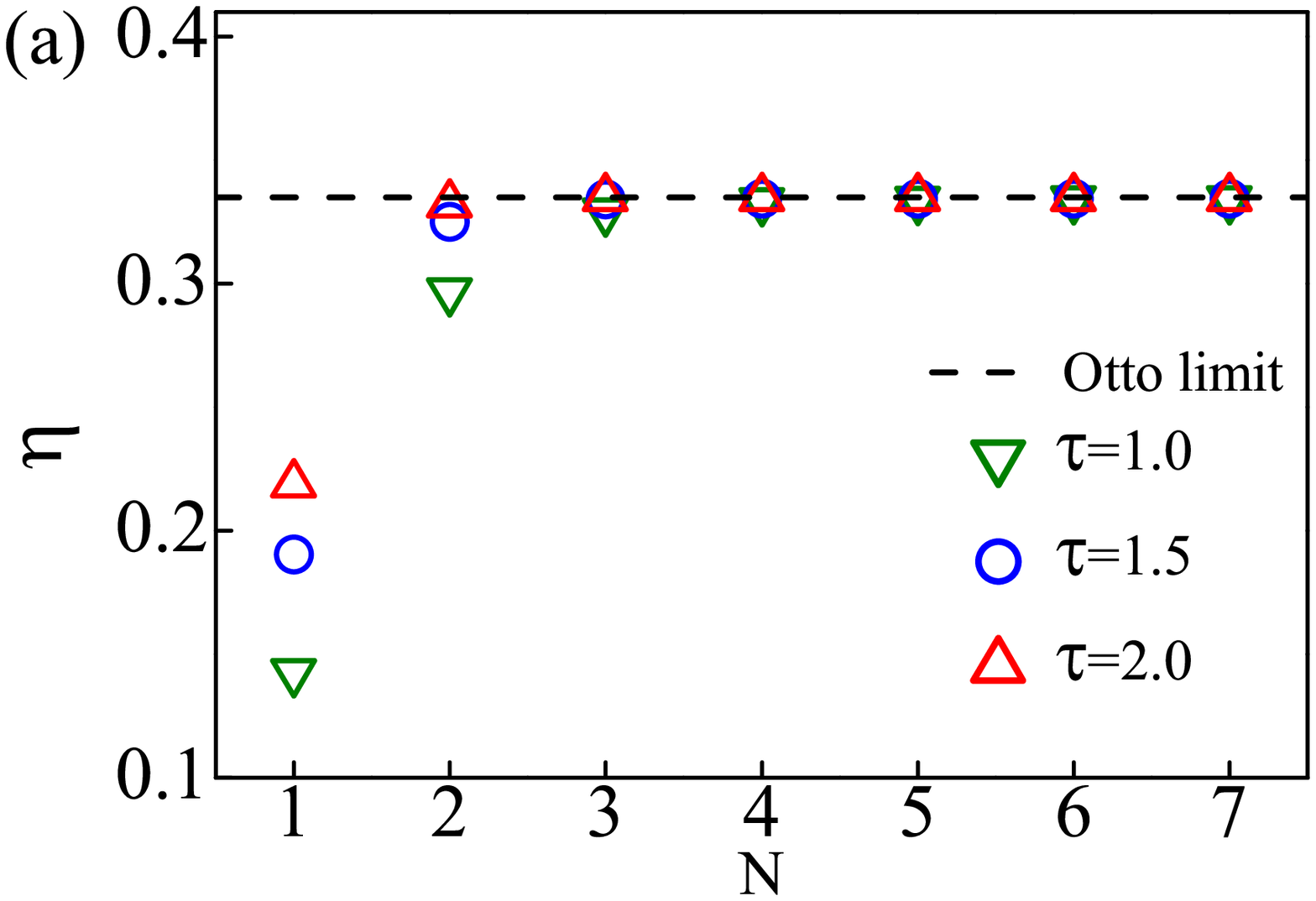}

\includegraphics[width=0.6\columnwidth]{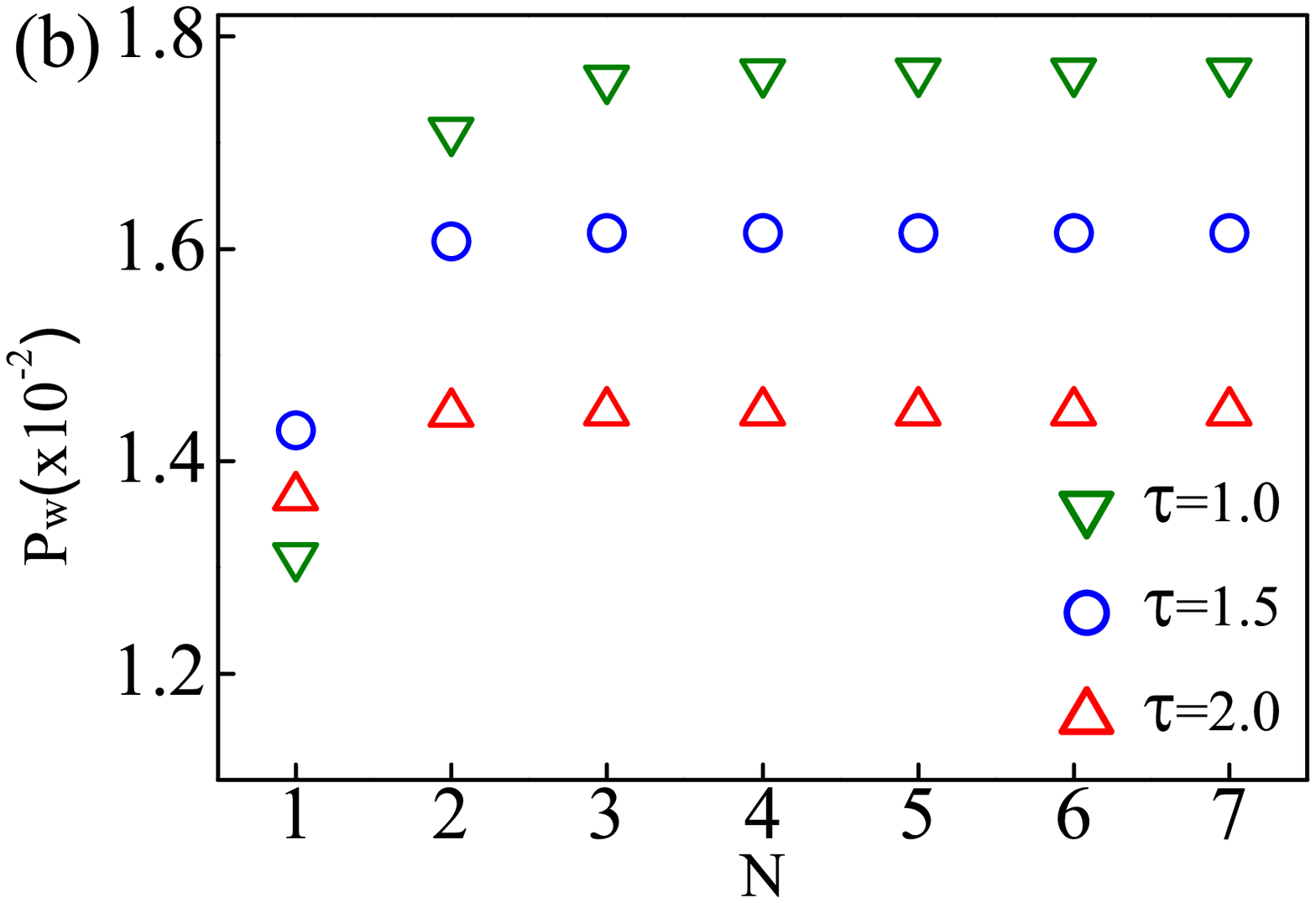}

\caption{The efficiency $\eta$ (a) and power $P_{W}$ (b) versus the number $N$ of the Otto cycles starting from the ground state. The
single-segment time $\tau$ takes $1.0$, $1.5$
and $2.0$. The temperature and frequency parameters are the same
as those in Fig. \ref{fig2}. The power is expressed
in relative unit of energy and time. \label{fig3}}
\end{figure}

Firstly, we set  the ground state of the oscillator as the initial state. The time evolution of the internal energy $U$ is shown in Fig. \ref{fig2}. The thermal engine reaches cyclostationary state after several periods.
With the number $N$ of the Otto cycle increasing, the quantum machine
gradually approaches periodically steady. For instance, the number
of particles on each state reach constant values at points $A$, $B$,
$C$, $D$, as shown the right of the separator in Fig. \ref{fig2}.
In addition, we obtain the evolution of efficiency at different working duration $\tau$ as the illustration in Fig. \ref{fig3}(a).
The Otto engine reaches periodic stability after several cycles. The longer the evolution time lasts, the closer $P_{n}^{A}$ is to
$P_{n}^{A'}$, which means that the efficiency $\eta$ is approaching
$\eta_{O}$, see Eq. \eqref{eqeta-1}. In the non-steady
cycle, the longer the working duration $\tau$ is, the higher the
efficiency $\eta$ is. Because the QOHE starts from the ground state,
$Q_{in}$ in the first-stroke is much larger than that in subsequent
cycles. Consequently, the efficiency is lower than the Otto limit
in the first stage. The efficiency $\eta$ of the QOHE could reach a stable
value after several cycles.  It is
worth noting that $P_{n}^{A'}=P_{n}^{A}$ holds only when the Otto
cycle reaches a steady cyclical state and the efficiency $\eta$ in
 Eq. \eqref{eqeta-1} reaches the Otto limit $\eta_{O}$. On the
other hand, as long as the probability of $P_{n}^{A}$ is different
from $P_{n}^{A'}$, the efficiency deviates from the Otto limit $\eta_{O}$.

The power output $P_{W}$ is another indicator to judge the performance
of the QOHE. It could be expressed as
\begin{equation}
P_{W}=W_{eff}/4\tau,\label{eqPw}
\end{equation}
where $4\tau$ is total time of an Otto cycle. The time evolution of power is illustrated in Fig. \ref{fig3}(b). 
During the initial
several cycles, the power is lower than the value in the stable cyclic
state. The working substance does not warm up to the optimal working
order in the beginning few cycles. The power is also lower according
to the Eqs. \eqref{eqWeff} and \eqref{eqPw}. The power reaches the maximum
as long as the cycle reaches a fixed periodic state. In addition,
the shorter the working cycle duration is, the higher the output power
is at the regular stable states, which is similar to that of classic
heat engines.

\begin{figure}
\includegraphics[width=0.58\columnwidth]{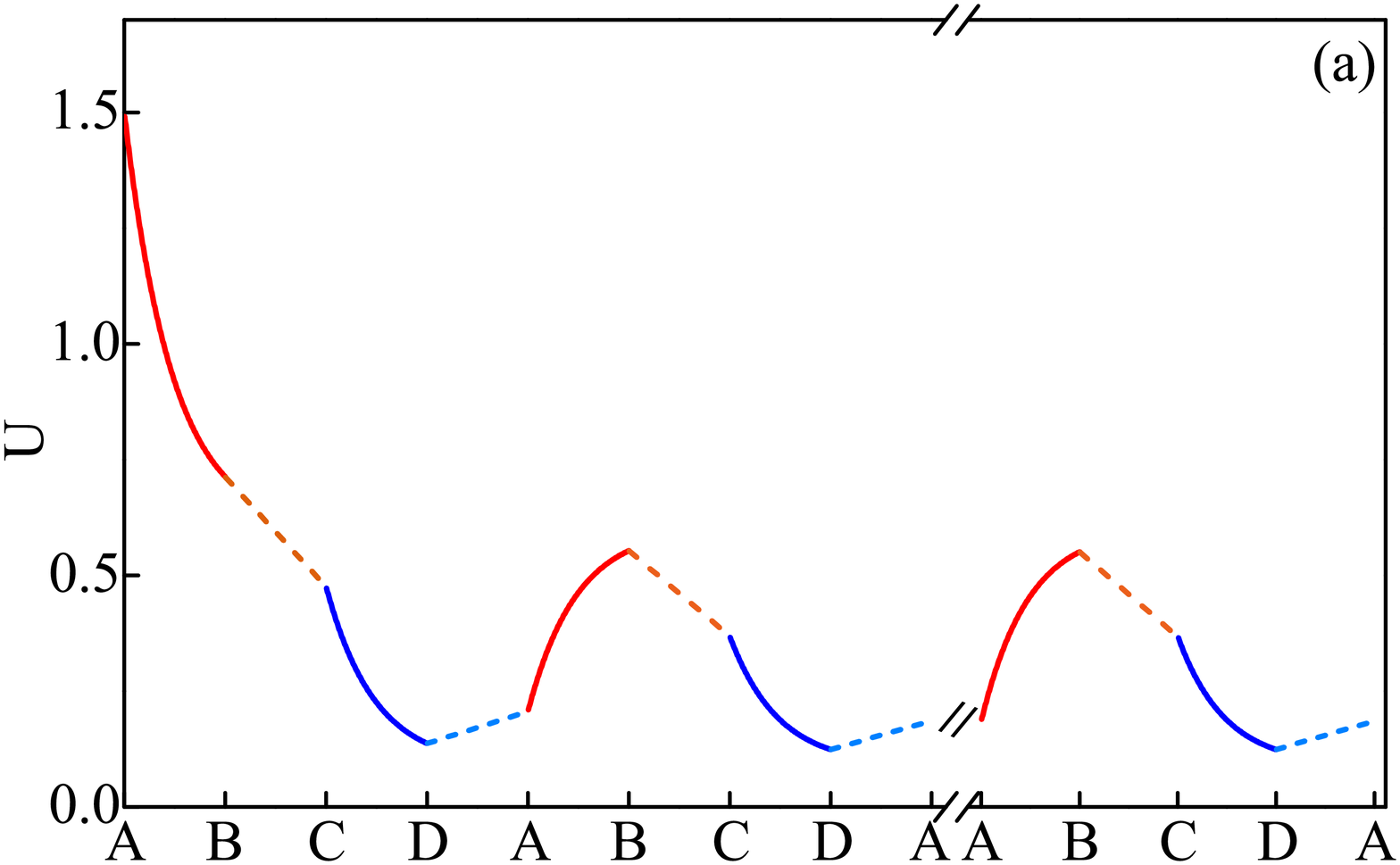}

\includegraphics[width=0.58\columnwidth]{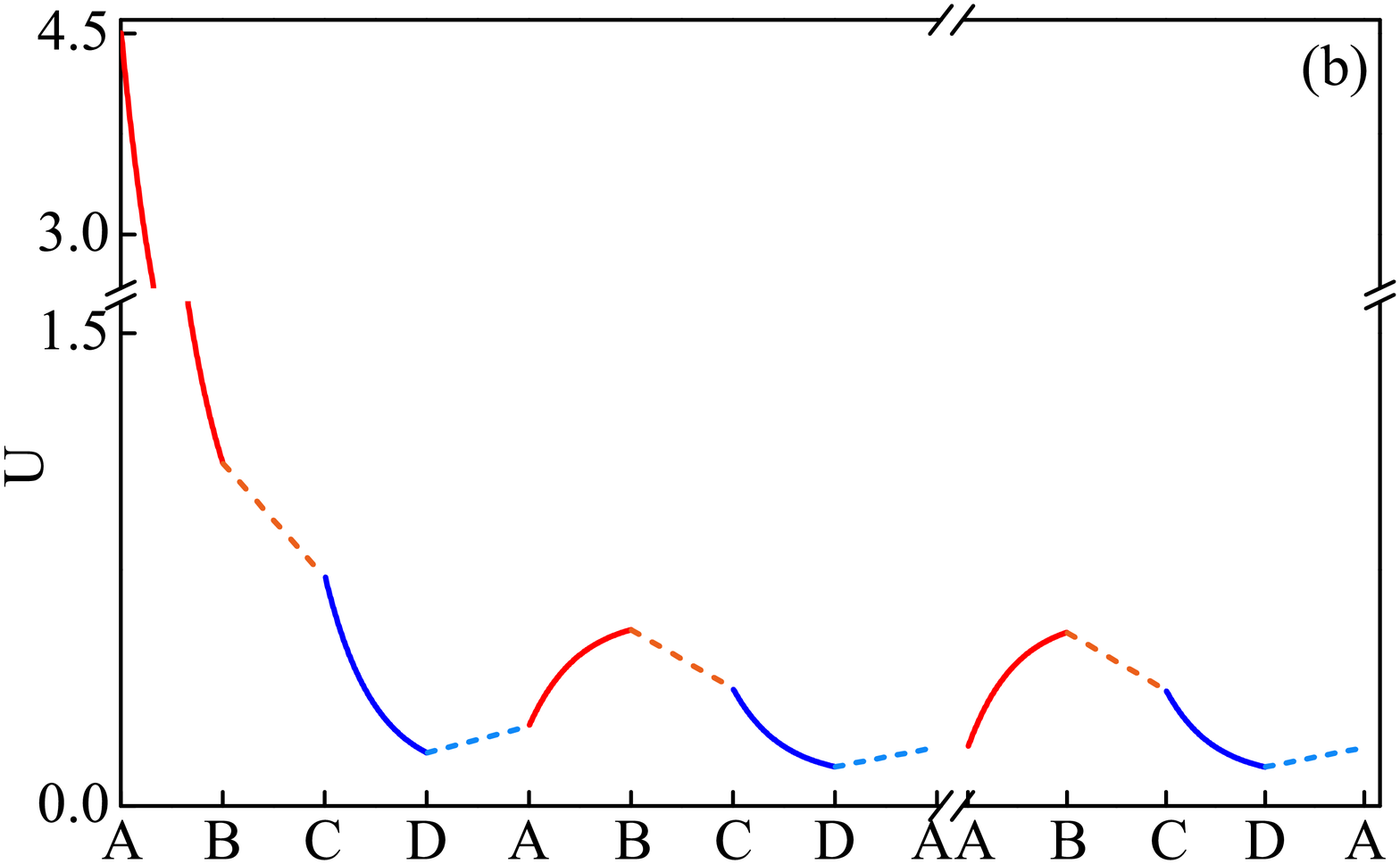}

\includegraphics[width=0.58\columnwidth]{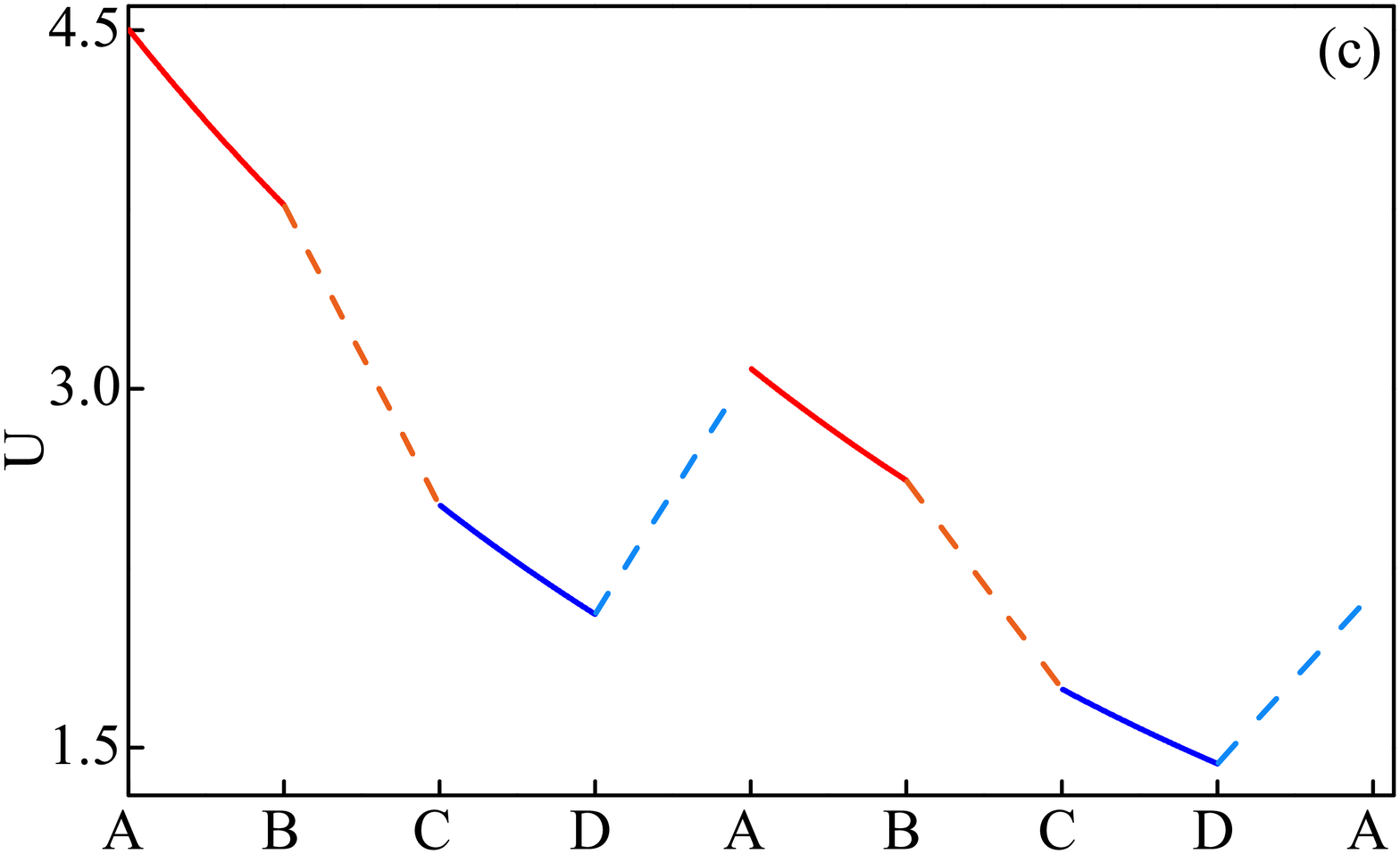}

\caption{The time evolutions of the energy $U$ (a) starting
from a initial state with the equal probability distribution on the
lower three energy levels. The energy $U$ (b) and (c) as the function of the time starting from the coherent state, and the relaxation time $(2\Gamma_0)^{-1}$ in the latter is ten times longer than that in the former to slow down the decoherence by environments. The parameters of temperature $T_{c,h}$,
frequency $\omega_{c,h}$ and the duration of single-stroke $\tau$
are the same as those in previous case starting from the ground state.
The red and blue solid lines represent endothermic and exothermic
strokes, respectively. The orange and light-blue dashed lines represent
the adiabatic working strokes.  \label{fig4}}
\end{figure}

Secondly, we set the initial state with the equal probability distribution
on the three lower energy levels of the oscillator. All the other
parameters such as the temperatures and frequencies are the same as
the previous case. The energy of this initial state
is much higher than that in the case starting from the ground state.
On account of the speedy redistribution of the probabilities among
the high energy levels in the initial stage, the energy quickly decays at the beginning
of the evolution and gradually tends to a periodic stabilization,
as shown in Fig. \ref{fig4}(a). In the starting periods,
the energy of the QOHE even releases into the hot bath in the form
of heat during the $\tau_{AB}$ or $Q_{in}<0$. 
\begin{figure}
\includegraphics[width=0.48\columnwidth]{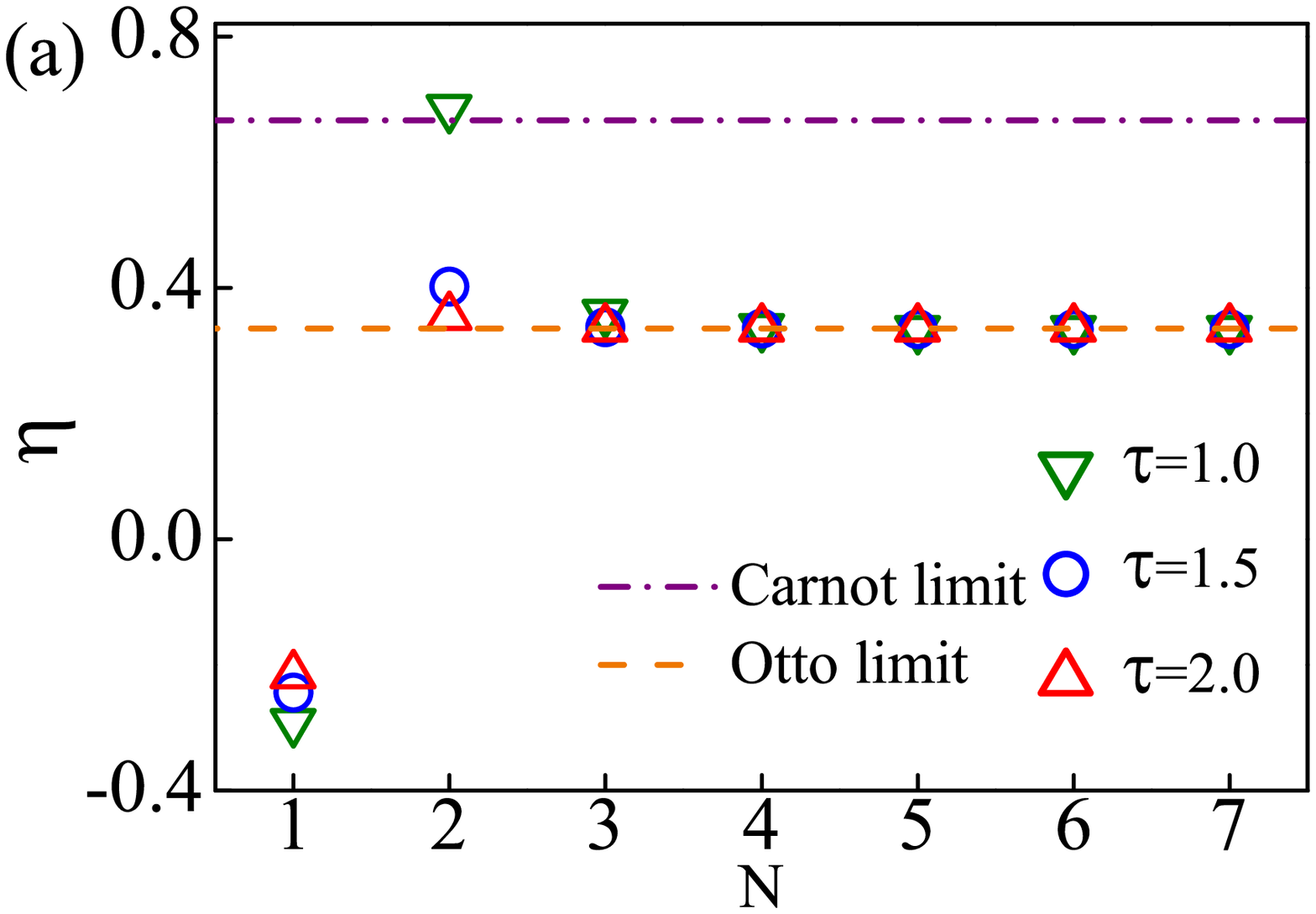} \includegraphics[width=0.48\columnwidth]{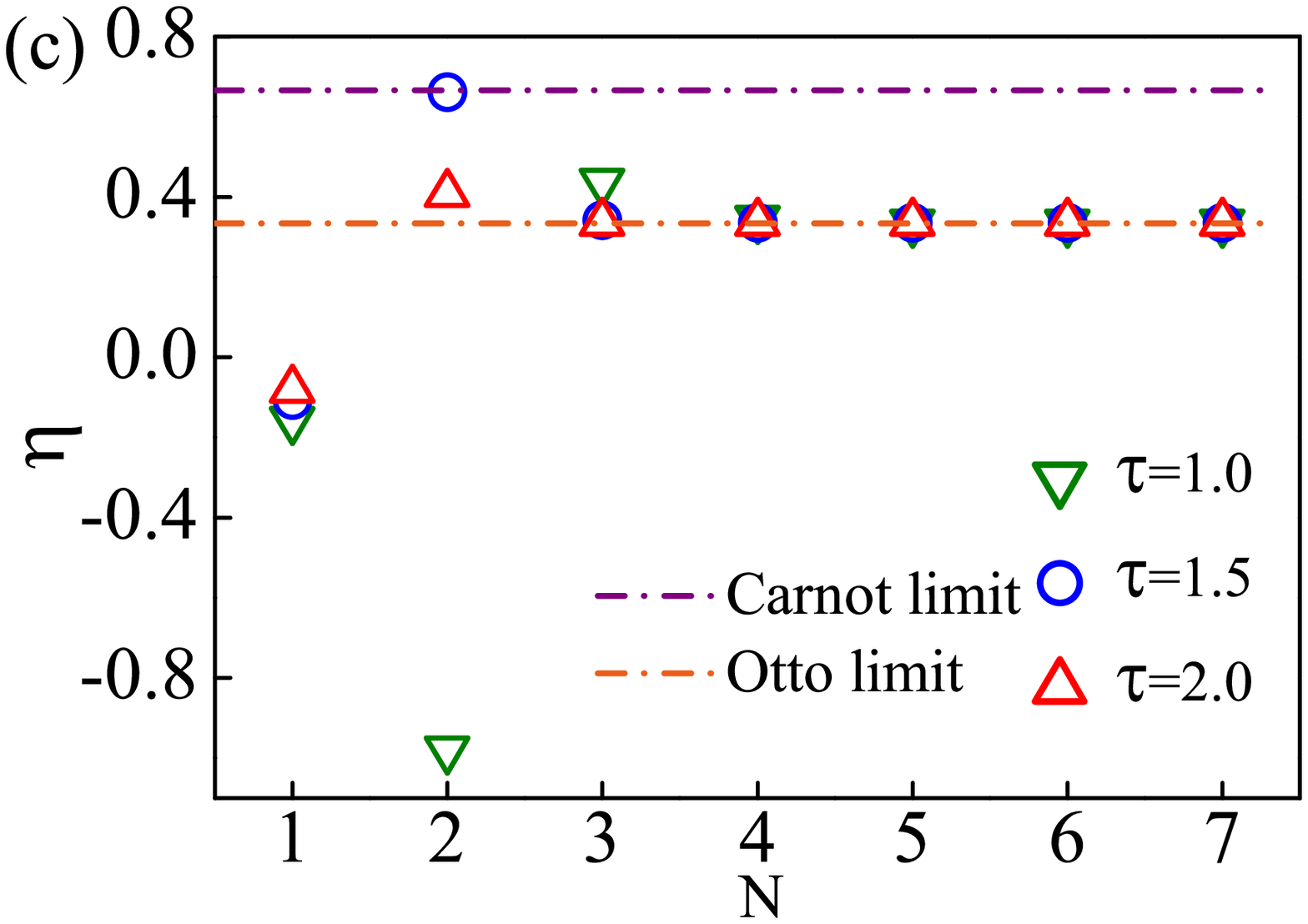}

\includegraphics[width=0.48\columnwidth]{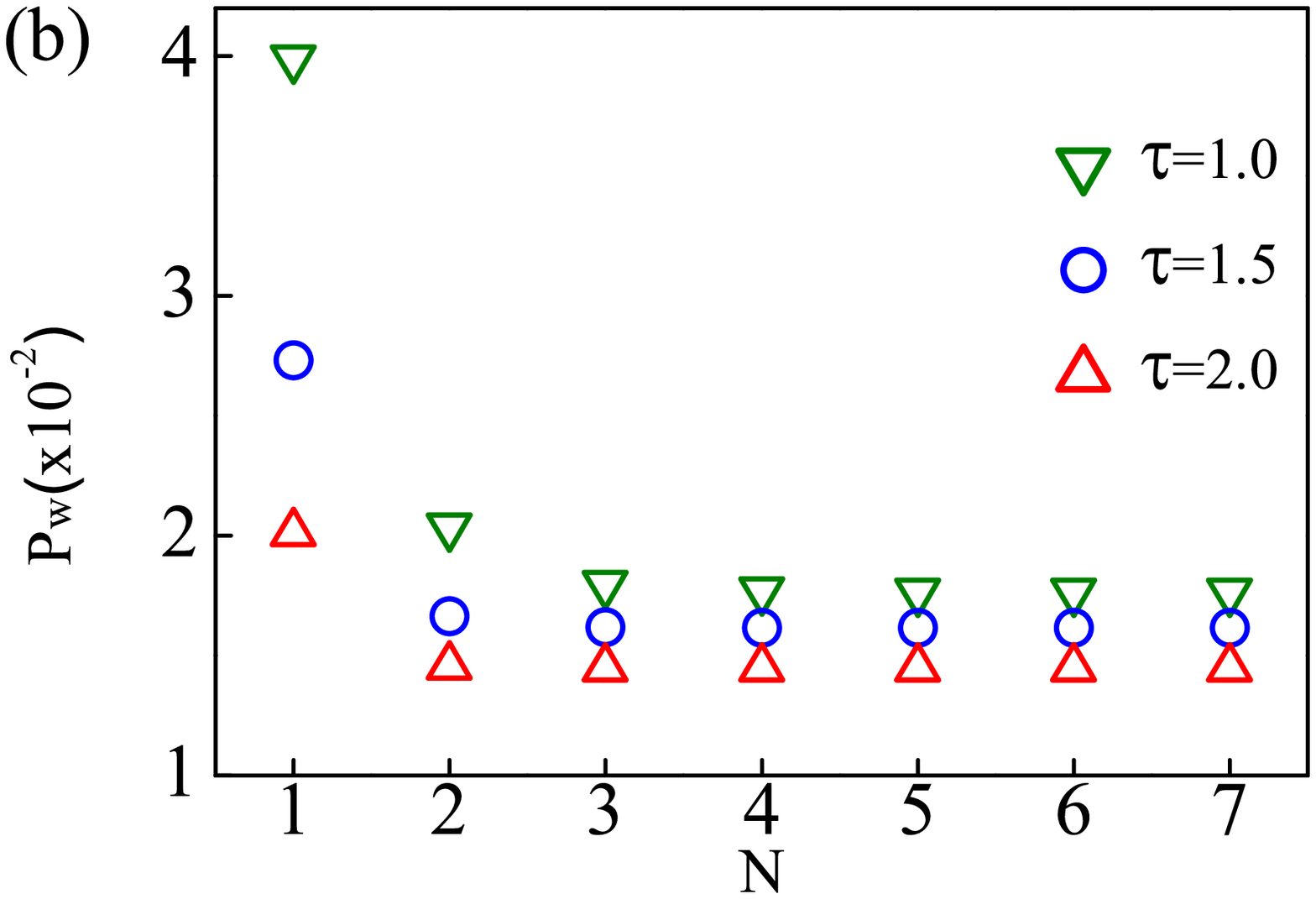} \includegraphics[width=0.48\columnwidth]{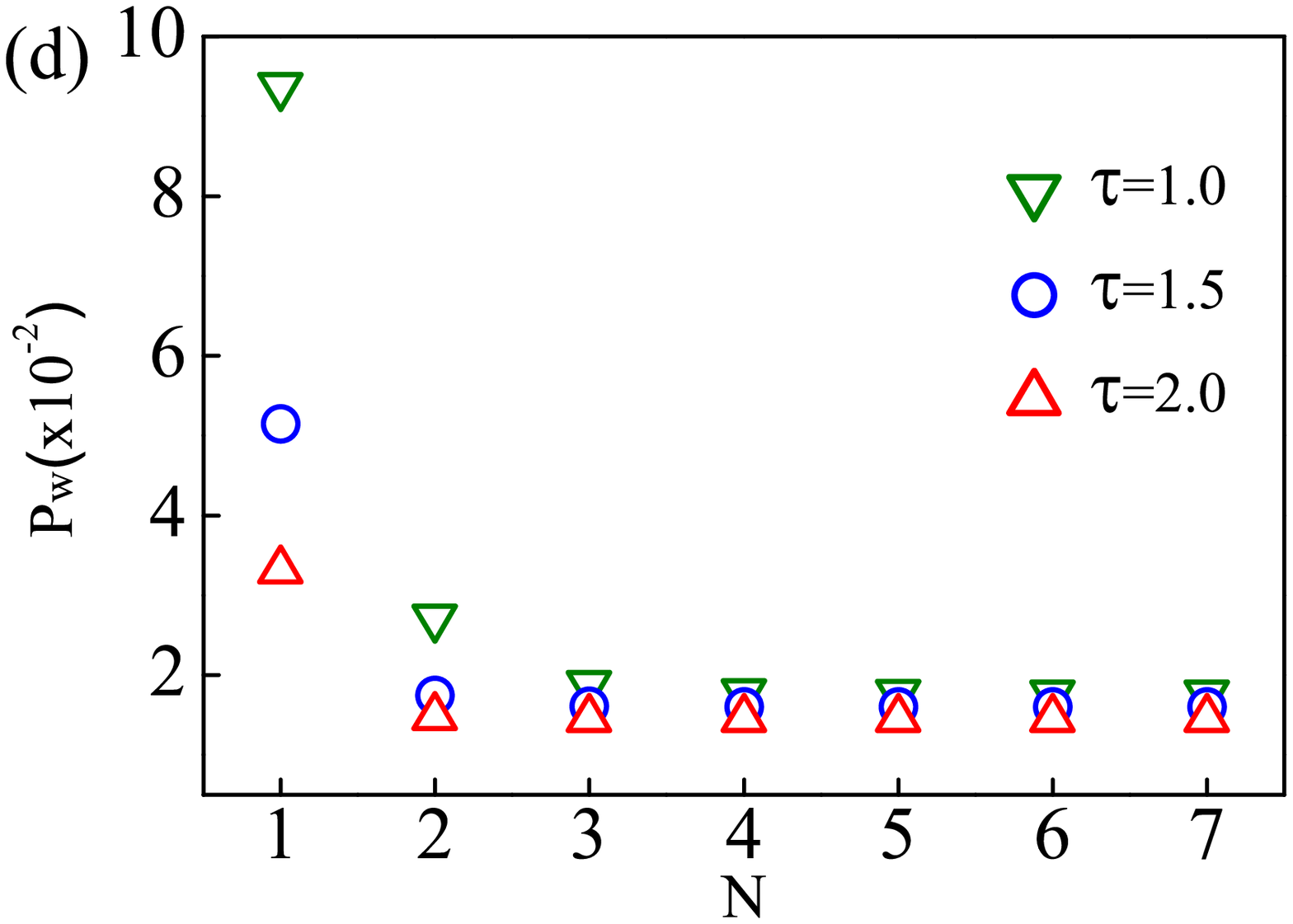}

\caption{The efficiency $\eta$ (a) and power $P_{W}$ (b) versus the number
of cycles $N$ with the equal probability distribution on the three lower
energy levels as the initial state. The efficiency $\eta$ (c) and power $P_{W}$ (d) versus the number
of cycles $N$ starting from the coherent state. The single-segment time $\tau$
takes different values. The temperature and frequency parameters are
the same as those in Fig. \ref{fig2}, respectively. \label{fig5}}
\end{figure}

Fig. \ref{fig5}(a) shows the time evolution of the efficiency.
In the beginning stages, one finds the negative efficiency value due
to the negative $Q_{in}$ and hence negative efficiency $\eta=W_{eff}/Q_{in}$.
The $Q_{in}$ gradually approaches zero and then becomes positive.
It is worth noting that in the $2$nd period, the efficiency even
exceed the Carnot or Otto limit. This does not means that the thermodynamic
laws are violated, and it is attributed to the contribution from the
energy stored in the initial state. Consequently, it provides a route
to improve the manipulations of the QOHE by external control, for example,
populating the high levels of the QHE by light or vibration pump.
In the stable periodic state of the QOHE, the efficiency returns to
the Otto limit $\eta_{O}$. The diagram of power evolution is shown 
in Fig. \ref{fig5}(b). The power decays from a higher
value to a constant. As the Otto cycle approaches a cyclostationary
state, the time evolutions of $U$, $\eta$ and $P_{W}$ are
all the same as those starting from the ground state. 

The enhancement of power by pumping potentially is of importance in realistic applications. For example, in a given environment, a heat engine works with rated power. If we come across a local obstacle that need larger power, then we need a substitute of the working engine. With the light or vibration pump, we could enhance the power beyond the rated power of the engine without replacing the engine.

Finally, we start from the coherent state with the occupation probability following the function $P_{n}\sim e^{-\left[\left(n-n{'}\right){\hbar\omega/k_{B}T}\right]^{2}}$ with $n$ for $n$th energy level and  $n{'}$ for the central level. However, once system-bath interaction is turned on, coherence loses quickly. The heat machine reaches a thermal equilibrium state within several cycles. The time evolution of the internal energy, efficiency and power is similar to those starting from the initial state with the equal probability distribution on the three lower energy levels, as shown in Fig. \ref{fig4}(b) and Fig. \ref{fig5}(c, d). To slow down the environment induced decoherence, we prolong the relaxation time $(2\Gamma_{0})^{-1}$ ten times longer, then the deformation of the wave packet with time could be reduced and the coherent state is more like a state of a classical oscillator at the beginning of the evolution. The energy increases or decreases almost linearly, as shown in the Fig. \ref{fig4}(c). The efficiency and power are strongly reduced (not shown).

\begin{figure}
\includegraphics[width=0.6\columnwidth]{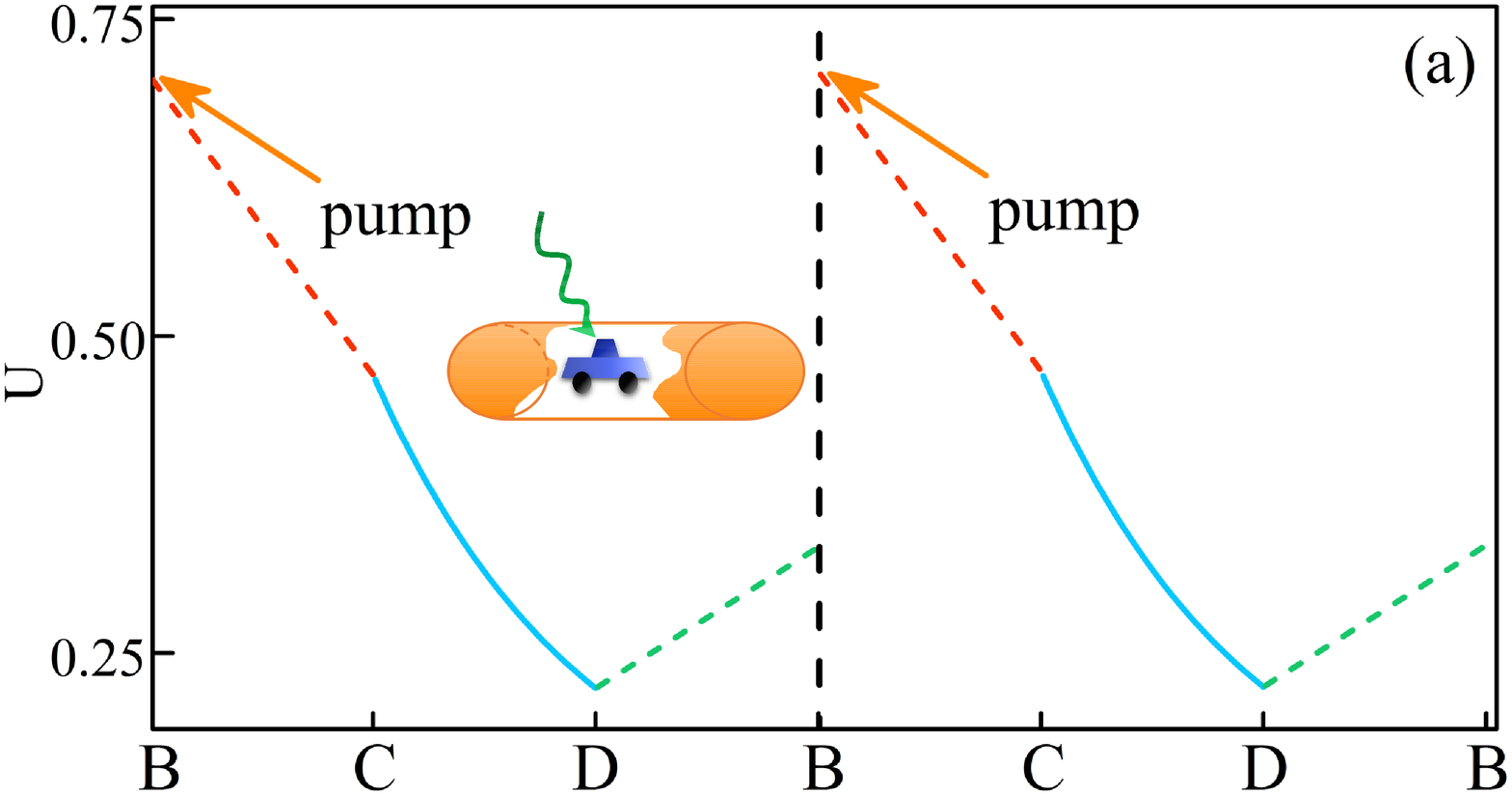}

\includegraphics[width=0.6\columnwidth]{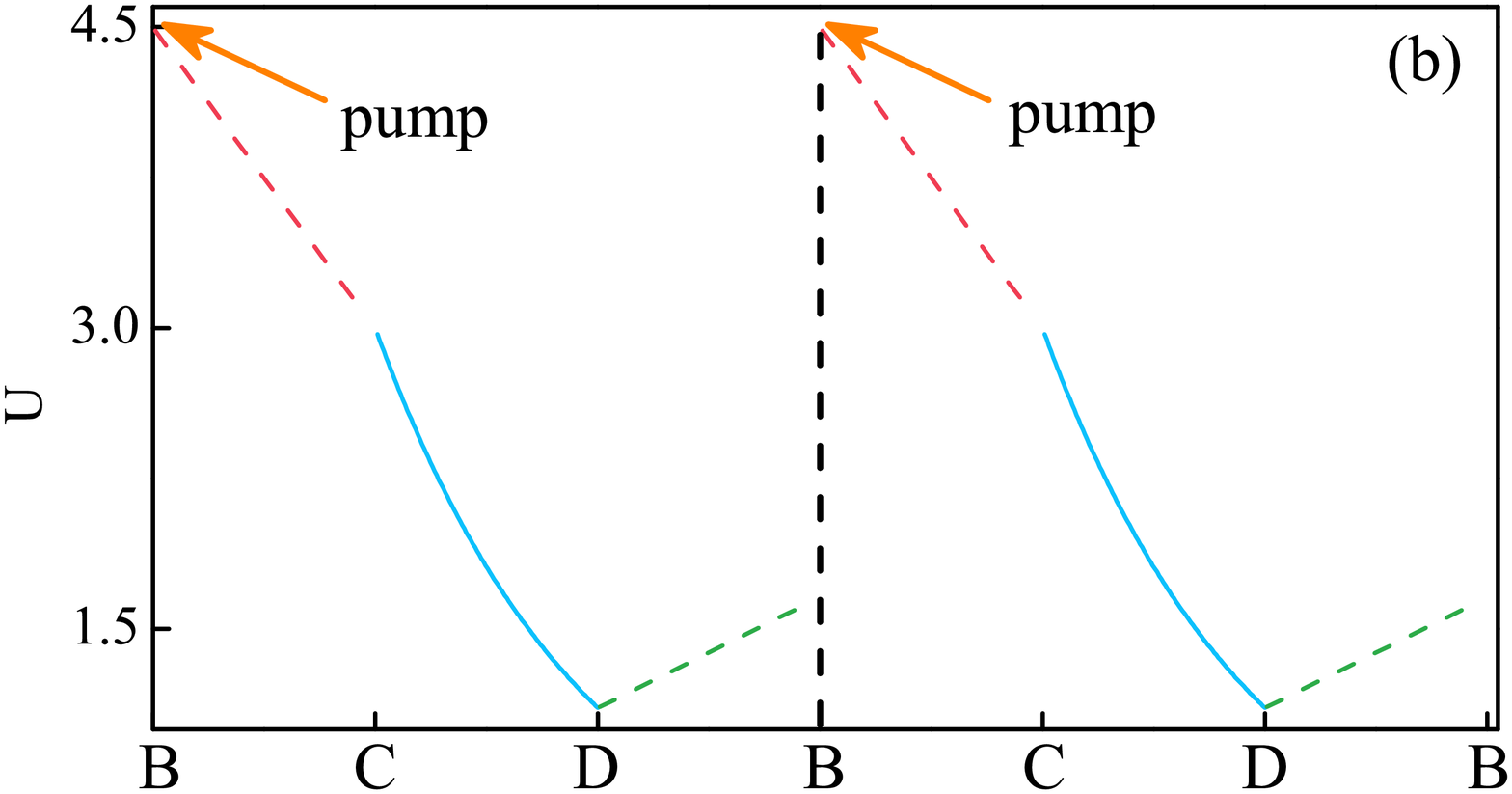}

\caption{The time evolution of the energy $U$ under periodically pumping in
a single heat reservoir. First energy level excited state (a) and coherent state (b). Here, single-stage duration $\tau=1.0$
and the thermal source at fixed $k_{B}T_{c}=0.4$. The frequencies
of harmonic oscillators are the same as the previous cases. The red
and green dashed lines show the segments of isentropic working; the light-blue
solid lines show the segments of isochoric heat release. The pumping
operation is performed at the beginning of each cycle as indicated 
by the orange arrows in all pictures. The pump duration $\tau_{AB}$ is ignored for it is much shorter than $\tau$. \label{fig6}}
\end{figure}

\subsection*{B. QHE with Single-bath \label{subsec:B.Single-bath-QHE}}

Heat engine always operates in two heat reservoirs. However, two baths
with different temperature are difficult to realize in microenvironments,
such as organism and nanostructures. Since the initial state affects the
time evolution of the power, internal energy and the efficiency
at least in the early QHE cycles, we could periodically pump the QHE
and operate it more easily. Therefore, here, we propose to pump
the quantum engine with a light or vibration pump and convert the
pump energy into the mechanical work in a single heat reservoir or
in a normal temperature environment. In other words, the energy provided
by pump substitutes the heat absorption from the hot bath in the Otto
cycles, and then the quantum engine operates in a single heat bath
with the temperature $T_{c}$. The engine is regularly pumped at the beginning
of the cycles to initialize the occupation probability distribution.
Since the pump duration $\tau_{AB}$ is extremely short, comparing
with the time segments of $\tau_{BC}$, $\tau_{CD}$ and $\tau_{DB}$,
we neglect $\tau_{AB}$ in our figures in this model. The evolution
process of the internal energy is shown in Fig. \ref{fig6}(a). The QOHE is periodically pumped from the ground state to the first
excited state at beginning of each cycle with the pump energy $Q_{pump}$. This engine actually converts the pump energy into heat and mechanical
work. The heat release process is necessary because
the pump energy is impossible to be converted into work completely
without the heat dissipation according to the second law of thermodynamics.
During the adiabatic strokes, the processes of the work input and
output are similar to the compression and expansion of classical heat
engine. The Otto cycle can be redivided into four processes: pump,
expansion, heat release and compression. The efficiency is redefined
as the ratio of the net work output to the pump energy, i.e., $\eta=\left(W_{out}+W_{in}\right)/Q_{pump}$.

\begin{figure}
\includegraphics[width=0.6\columnwidth]{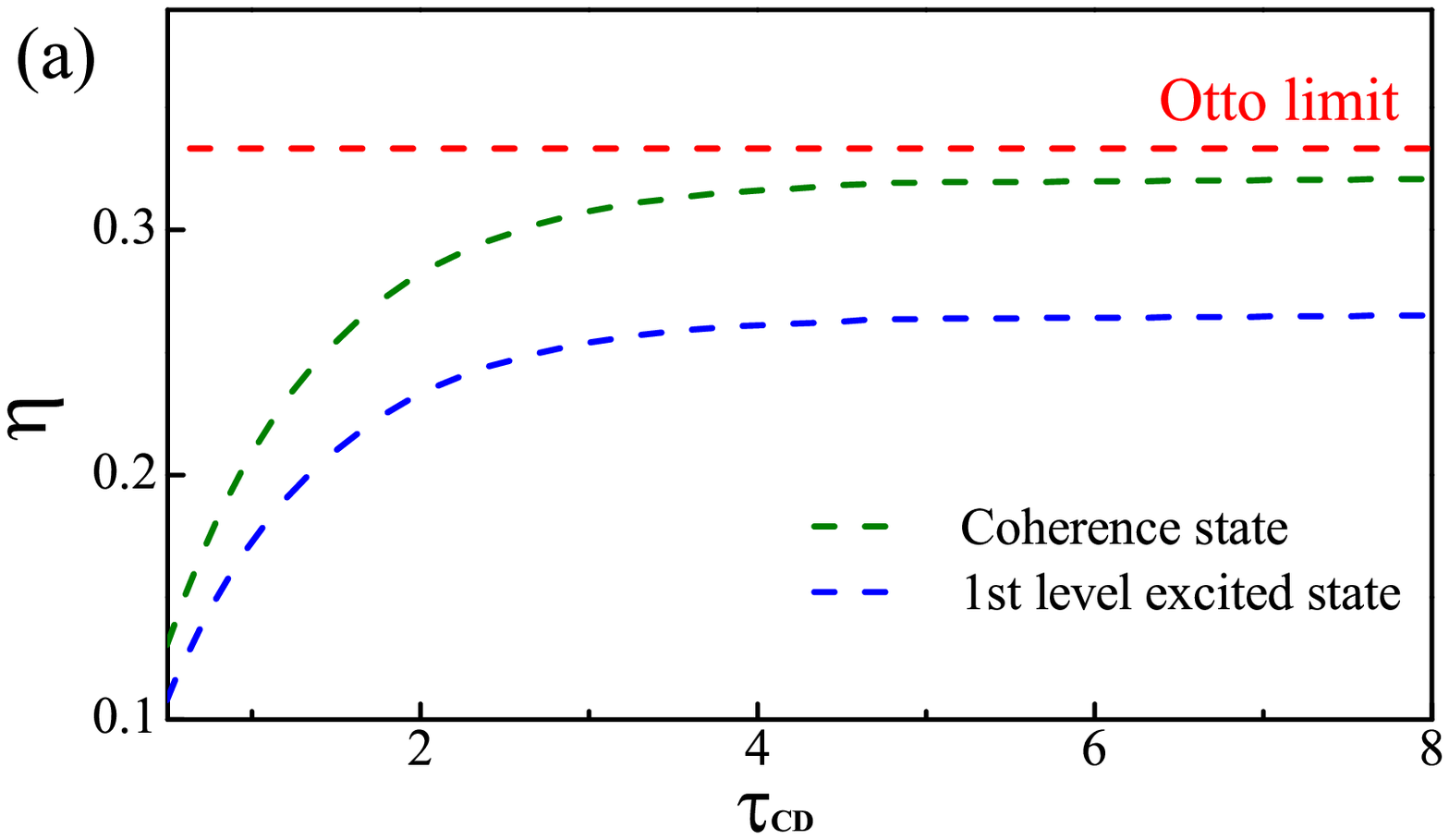}

\includegraphics[width=0.6\columnwidth]{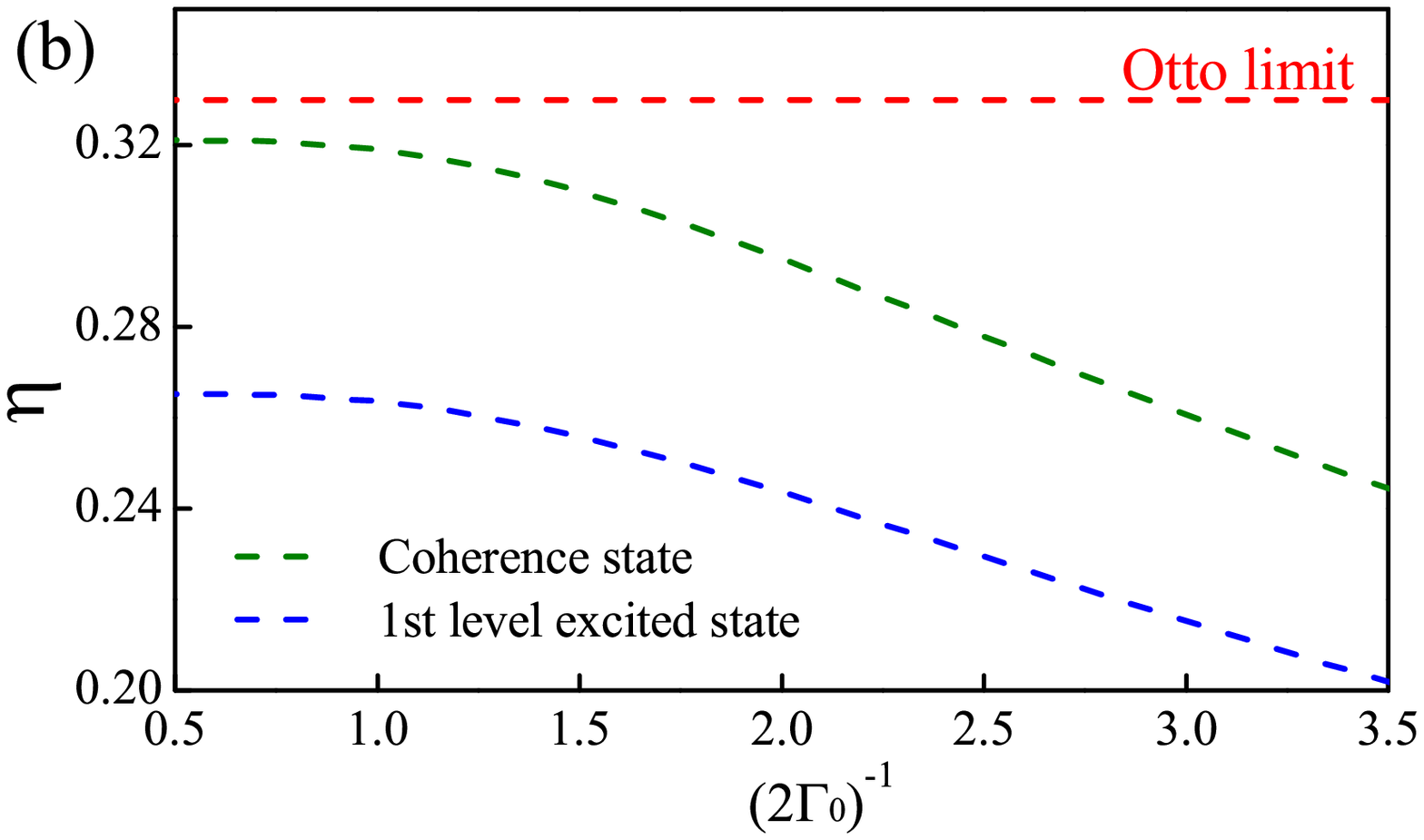}

\caption{Dependence of efficiency on thermalization time and relaxation time. (a) efficiency vs variant system-environment interaction time $\tau_{CD}$ with the work time $\tau_{BC}=\tau_{DB}=1$, the relaxation time constant $\left(2\Gamma_0\right)^{-1}=1$. (b) efficiency vs the relaxation time parameter $\left(2\Gamma_0\right)^{-1}$. The time interval for all stages are $\tau_{BC}=\tau_{DB}=1,~\tau_{CD}=5$, respectively. The frequencies of harmonic oscillators and temperature are the same as the previous cases. 
\label{fig7}}
\end{figure}

Actually, the pump method could easily prepare variants of initial states. We set the coherent state as the initial state to start engine. The internal energy as a function of the working time is illustrated in Fig. \ref{fig6}(b). The time evolution is similar to that of the first level excited state. In Fig. \ref{fig7}, starting from the coherent state, one finds that the efficiency is significantly improved comparing with the case starting from the first level excited state, . Since coherence plays an important role in improving the performance of QHEs \cite{Camati2020,Guff,Dodonov}, we further study the state coherence effects. The quantum coherence is weakened as the system interacting with environments \cite{Gong}. The longer the time lasts, the less coherence is left and the less input work is needed to start the next cycle. In this pump-driven machine, the system environment interaction takes place only in the heat exchange phase $\tau_{CD}$. We study the efficiency as the function of the time variable $\tau_{CD}$. In Fig. \ref{fig7}(a), the efficiency is improved with the time $\tau_{CD}$ increasing until the time ratio $\tau_{BC}/\tau_{CD}\sim 0.2$, where the efficiency reaches the upper bound. According to the results in Fig. \ref{fig7}(a), the efficiency reaching its maximum implies the thermalization is completely carried out. Another significant factor is the relaxation constant $\Gamma_0$, which reflects the speed of relaxation. Here, we set the relaxation time $\left(2\Gamma_0\right)^{-1}$ as the variable parameter. As $\left(2\Gamma_0\right)^{-1}$ increases, the complete thermalization stroke takes a longer time, and as shown in Fig. \ref{fig7}(b), the efficiency gradually deviates from the Otto limit. Therefore, we could intervene into the initial state by periodic pumping to improve the performance of heat engines.

The pump-driven engine has important potential applications. In microenvironments, such as living organisms and nanostructures, it is almost impossible to provide two baths with large temperature difference. The single bath engine driven by light pumping outside is a quite suitable candidate to work in microenvironments. For instance, under the controllable light irradiation, atomic dimers, quantum dots or magnetic clusters absorb photons and then the chemical bonds prolong and perform mechanical motion \cite{Hubner2014,Benenti2017}. Equipped with this kind of engines, biological machines or molecule cars could be used to identify and destroy lesions \cite{Novotny}. For example, to purge the clots in blood vessel, a nanorobot could be sent to the location of the clots to deliver drugs or directly dredge the clogged blood vessel. The nanorobot equipped with the pump-driven engine could work under external pumping control without a cable or a battery. 

\begin{figure}
\includegraphics[width=0.55\columnwidth]{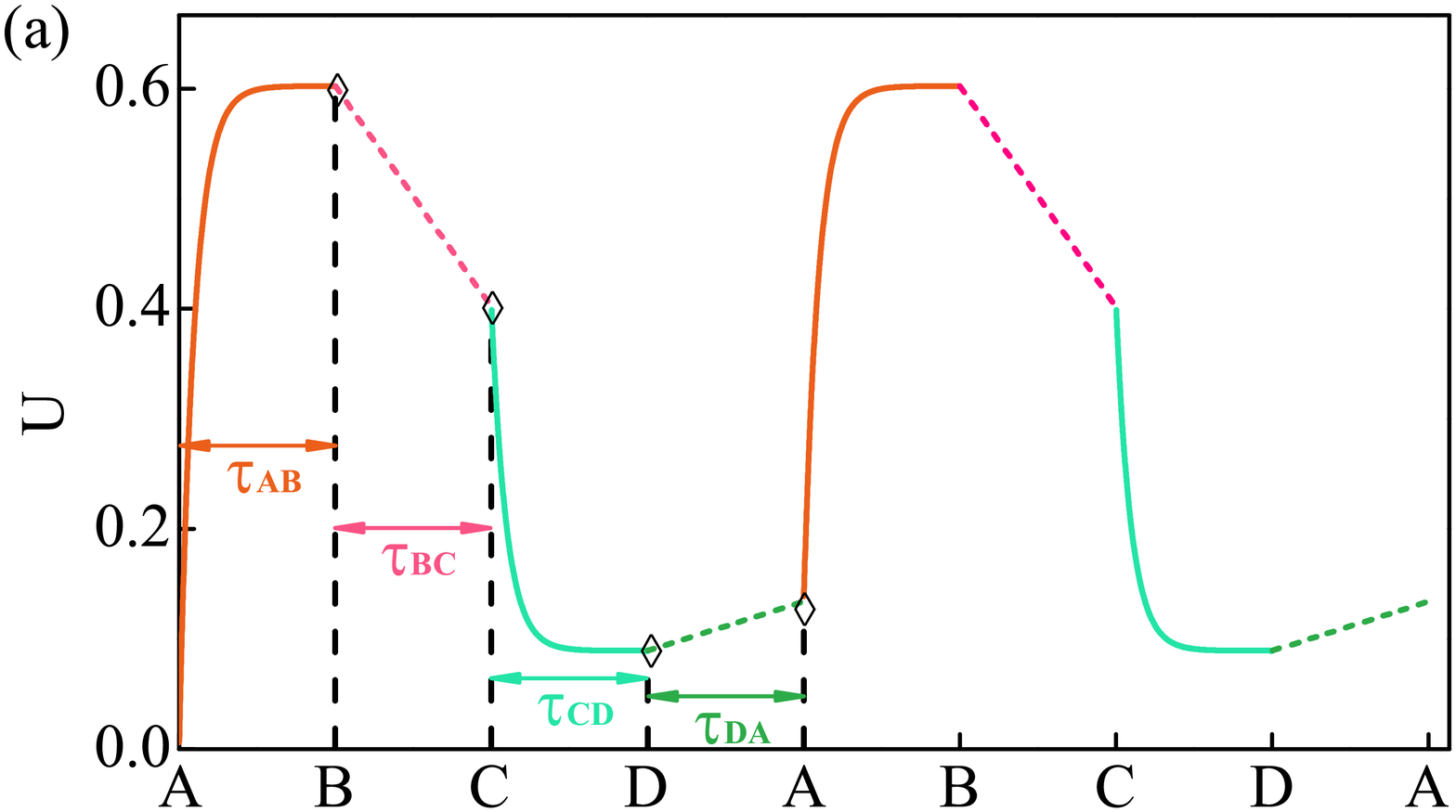}
\includegraphics[width=0.55\columnwidth]{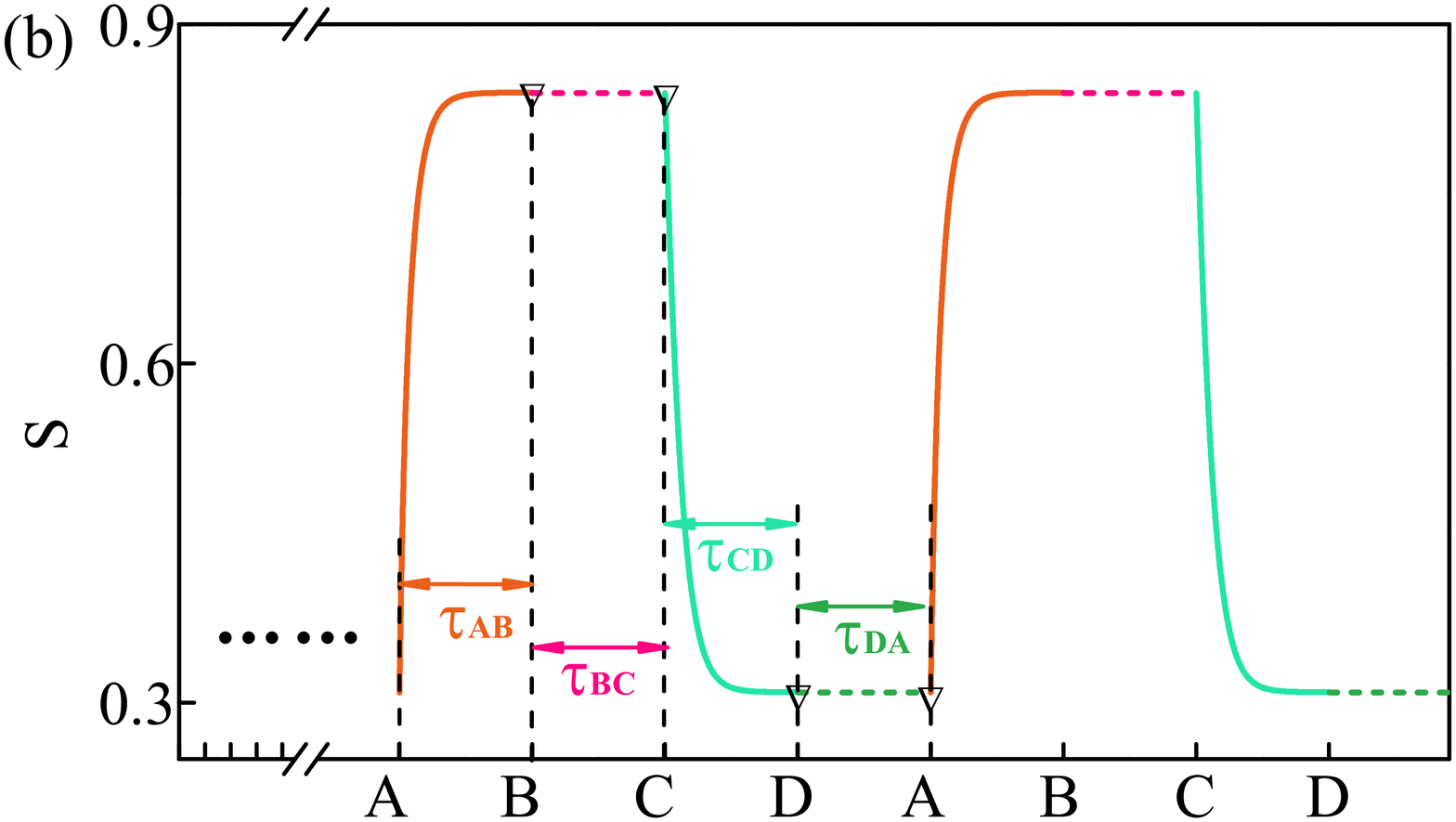}
\caption{The time evolutions of the energy $U$ (a) with
the QOHE reaching thermal balance with the baths. $\hbar\omega_{c}=1.0$
and $\hbar\omega_{h}=1.5$. (b) Entropy evolution diagram after the quantum state coherence disappears completely. We set the working time of single-step
 $\tau=10.0$. The temperature parameters of cold bath and hot bath
are $k_{B}T_{c}=0.4$ and $k_{B}T_{h}=1.2$. The pink and green solid
lines display the isochoric heat exchange stages; The orange and light-green
dashed lines display the isentropic working stages. The hollow diamond
and triangle symbols represent analytical results, exactly locating
on the numerically calculated curves. \label{fig8}}
\end{figure}

\subsection*{C. Otto cycle with thermal balance\label{subsec:C.Steady-State-QHE}}

We have studied the performance of the QOHE without forming thermal
equilibrium with the heat baths in the above sections. In the following,
we prolong the duration of periodic cycle to make the machine reach
the thermal equilibrium with the baths in the heat exchange processes.
The harmonic oscillator's frequency and environmental conditions are
exactly the same as those in the previous sections.
When the QOHE reaches thermal balance with the heat reservoirs, the
population probabilities obey the Boltzmann distribution, $P_{n}\propto e^{-E_{n}/k_BT}$.
 During the heat-exchange
processes, the frequencies of the harmonic oscillators remain unchanged.
The population distribution function $P_{n}$ at the thermal balance
could be calculated analytically. The initial state also starts from the ground
state, and the cycle reaches stable in the first stage $\tau_{AB}$.
Only in the first cycle, the absorbed heat and efficiency are different
from those in the subsequent cycles. The time evolution of the entropy is shown in the Fig. \ref{fig8}(b). Since Eq. \eqref{eqS} fails to apply to the entropy calculation of the states with coherence. We omit the stating stages of the Otto cycles until the coherence completely is eliminated.
The orange and light-green dashed lines display the isentropic
working stages. 

Our numerical calculations of energy and entropy qualitatively agree well with the results obtained by the Lindblad methods \cite{Dodonov,Jong-Min2019}. In addition, the data points based on analytical calculation at the points $A,\ B,\ C,\ D$ also exactly locate on our numerically calculated curves, as shown in Fig. \ref{fig8}.

\begin{figure}
\includegraphics[width=0.55\columnwidth]{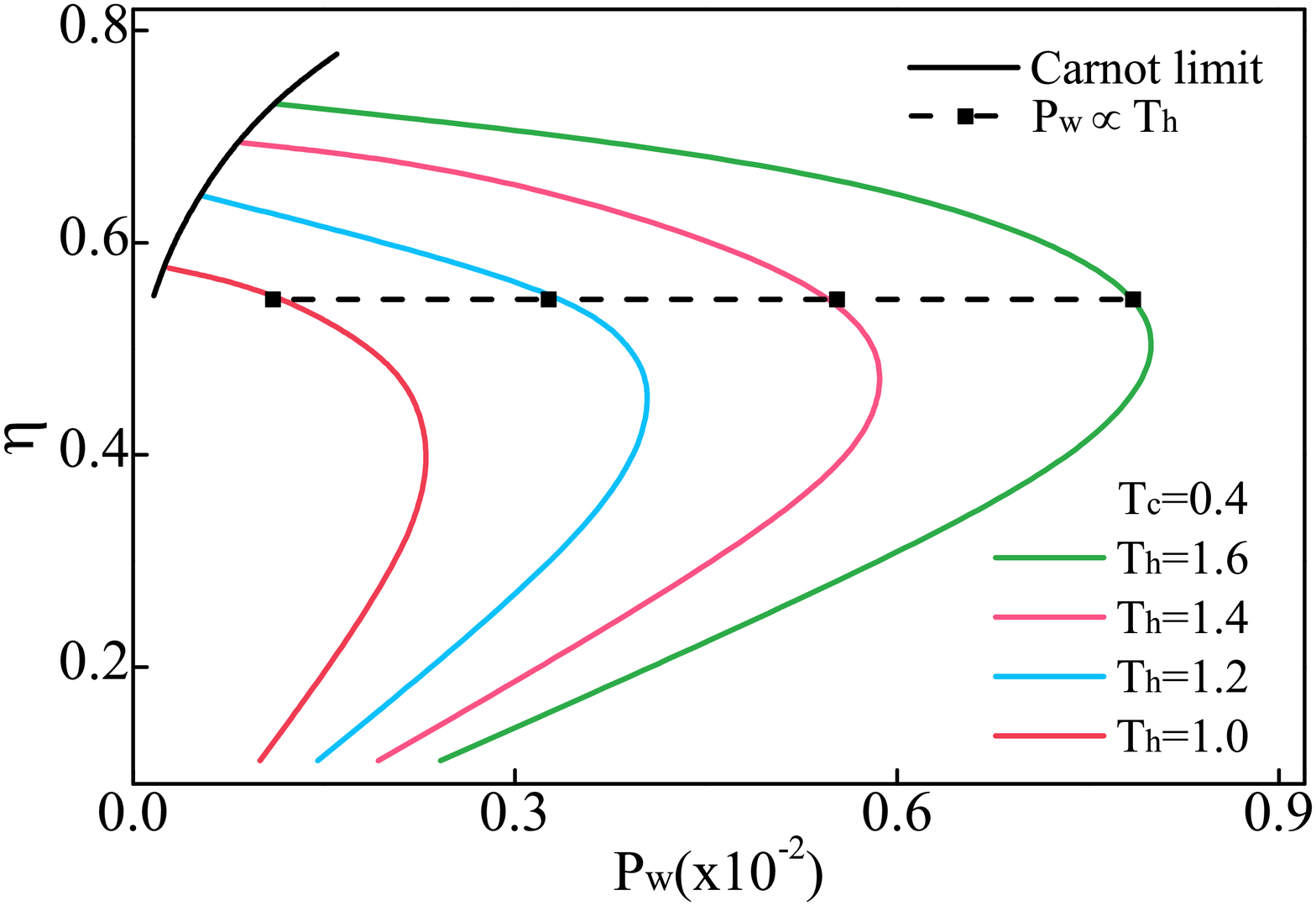}

\caption{The efficiency $\eta$ versus the effective power output $P_{W}$
of QOHE. Different colored lines represent the relationship between
efficiency and net output power at different $T_{h}$ by gradually
changing the frequency ratio $\omega_{c}/\omega_{h}$. The cold bath temperature $k_{B}T_{c}=0.4$. The black horizontal dashed line sketches
the power $P_{W}$ vs $T_{h}$ at a fixed $\eta$. Black solid line
indicates the Carnot limit, constraining the efficiency of the QOHE.\label{fig9}}
\end{figure}

QHEs often fail to maximize both the output power and the efficiency \cite{Abah2012}.
The balance between the power and efficiency is pursued. We fix the
cold bath temperature a constant value, and adjust the hot bath temperature
as well as the frequency ratio $\omega_{c}/\omega_{h}$. The relationship
between the efficiency versus the output power is shown in Fig.
\ref{fig9}. The power versus efficiency shows a quasi-parabolic
curve. It is impossible for both of them to reach the maximum. The
highest efficiency of a QOHE is limited by the Carnot limit $\eta_{C}=1-T_{c}/T_{h}$ \cite{Kieu2004,Kieu2006,Abah2014,Jong-Min2019}
with the ratio $\omega_{c}/\omega_{h}=T_{c}/T_{h}$. With the same
efficiency, the higher temperature of the hot bath  leads to the higher output
power, as shown by the black dashed line in Fig. \ref{fig9}.
At the fixed temperature ratio $T_{h}/T_{c}$, the efficiency could
be improved with the frequency ratio $\omega_{h}/\omega_{c}$ enhancement
until it reaches Carnot limit.

\section{conclusion\label{sec:conclusion}}

To conclude, we have probed the time evolution of the quantum Otto
cycle process and the performance of the QOHE with a single oscillator as the working substance. We calculated the time-dependent population
distribution of the oscillator's energy levels by solving the driven-dissipative
Schr\"{o}dinger equation and simulated the time evolution of the internal
energy and power and efficiency. We show that the different
initial states have different impacts on these quantities in the transient
period before the Otto cycle becomes periodical stable. In the transition
time, the efficiency and power differs from the corresponding values
in the stable Otto cycles. The efficiency even surpasses the Otto
limit and the Carnot limit and  the efficiency anomaly is attributed to the contribution
from the energy stored in the initial state. Therefore, we suggest that
the periodically pumping could strongly increase the rated power and also could replace the hot bath. Furthermore, we propose a novel quantum engine in ambient
condition to convert the pump energy into the mechanical work. Such
an engine could work in the microenvironments without a large temperature
difference, such as biological tissues in vivo. We expect that the
operational protocol presented here is applied to modeling a quantum
engine with the advantage of controllability. 

\textit{Acknowledgments}.$-$We are thankful to Jize Zhao for fruitful
discussions. This work is supported by the National Natural Science Foundation of China No.91750111.

\end{document}